\documentclass{PoS}
\usepackage[utf8]{inputenc}
\usepackage{bbold}
\usepackage{amsmath}

\usepackage{mcite}
\usepackage{cite}

\title{Results on the heavy-dense QCD phase diagram using complex Langevin}

\ShortTitle{Results on the heavy-dense QCD phase diagram using complex Langevin}

\author{Gert Aarts\\
Department of Physics, College of Science, Swansea University,\\
Swansea, SA2 8PP, United Kingdom\\
E-mail: \email{g.aarts@swan.ac.uk}}

\author{\speaker{Felipe Attanasio}%
\\
Department of Physics, College of Science, Swansea University,\\
Swansea, SA2 8PP, United Kingdom\\
E-mail: \email{pyfelipe@swan.ac.uk}}

\author{Benjamin J\"ager\\
Department of Physics, College of Science, Swansea University,\\
Swansea, SA2 8PP, United Kingdom\\
E-mail: \email{b.jaeger@swan.ac.uk}}

\author{D\'enes Sexty\\
Department of Physics, Bergische Universit\"{a}t Wuppertal,\\
Gaussstra{\ss}e 20, D-42119 Wuppertal, Germany\\
and Inst. for Theoretical Physics, E\"otv\"os University,\\
P\'azm\'any P. s\'et\'any 1/A, H-1117 Budapest, Hungary\\
E-mail: \email{sexty@uni-wuppertal.de}}

\abstract{Complex Langevin simulations have been able to successfully reproduce results from Monte Carlo methods in the region where the sign problem is mild and make predictions when it is exponentially hard. We present here our study of the QCD phase diagram and the boundary between the confined and deconfined phases in the limit of heavy and dense quarks (HDQCD) for 3 different lattice volumes. We also briefly discuss instabilities encountered in our simulations.}

\FullConference{34th annual International Symposium on Lattice Field Theory\\
		24-30 July 2016\\
		University of Southampton, UK}

\begin{document}

\section{Introduction}
\begin{figure}[h!]\center
	\includegraphics[scale=0.7]{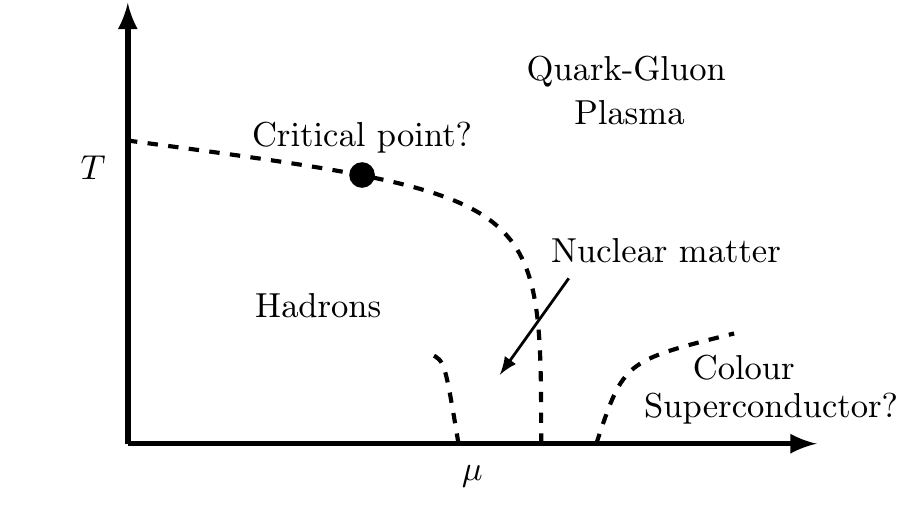}
	\vspace{-12pt}
	\caption{Sketch of the possible QCD phase diagram.}
	\label{fig.phase.diagram}
\end{figure}

For the past three decades countless non-perturbative results of QCD have been obtained using the lattice formulation, such as decay constants, hadron masses and properties of the thermal transition to the quark-gluon plasma at zero density.
However, a full picture of the QCD phase diagram at finite temperature and chemical potential, sketched in Figure \ref{fig.phase.diagram}, requires research at non-zero density as well.

A finite quark chemical potential $\mu \equiv \mu_q$ in the Euclidean path integral formulation leads to the so-called \textit{sign problem}, i.e., to a complex probability weight.
In some situations, where the sign problem is considered ``mild'', standard Monte-Carlo based methods, such as reweighting, and properties related to analyticity, like Taylor expansion, can still be applied \cite{deForcrand:2010ys}. However, these techniques cannot probe reliably regions where $\mu / T \gtrsim 1$ due to an exponentially hard overlap problem.

This difficulty is due to the quark determinant, which appears once the quark fields have been integrated out, since it is complex for real chemical potential $\mu$,
\begin{equation}
	\left[ \det M(U, \mu) \right]^* = \det M(U, -\mu^*)\,,
\end{equation}
where $U$ generically represents the gauge links.

We present here our results on the phase diagram of QCD in the heavy-dense approximation, to be explained below, obtained using complex Langevin simulations. We also discuss instabilities found in the simulations.
This contribution is based on \cite{Aarts:2016qrv}.

\section{Complex Langevin equation}
Stochastic quantisation \cite{Parisi:1980ys} is a procedure that allows quantum expectation values to be evaluated as averages over a stochastic process by evolving the dynamical variables in a fictitious time $\theta$ using a Langevin equation,
\begin{equation}
	U_{x\mu}(\theta + \varepsilon) = R_{x\mu} U _{x\mu}(\theta) \,, \quad R_{x\mu} = \exp\left[ i\lambda^a(\varepsilon -D^a_{x\mu}S + \sqrt{\varepsilon} \eta^a_{x\mu}) \right]\,.
\end{equation}
The gauge links are represented by $U_{x\mu}$, $\lambda^a$ are the Gell-Mann matrices, $\varepsilon$ is the stepsize and $\eta^a_{x\mu}$ are white noise fields satisfying
\begin{equation}
	\langle\eta^a_{x\mu}\rangle = 0\,, \quad \langle \eta^a_{x\mu} \eta^b_{y\nu} \rangle = 2 \delta^{ab} \delta_{xy} \delta_{\mu\nu}\,.
\end{equation}
The drift $-D^a_{x\mu} S$ is derived from the QCD action, $S = S_{YM} - \ln \det M$, and the gauge group derivative $D^a_{x\mu}$ is defined as
\begin{equation}
	D^a_{x\mu} f(U) = \left.\frac{\partial}{\partial \alpha} f(e^{i\alpha\lambda^a} U_{x\mu})\right|_{\alpha=0}\,.
\end{equation}
We have used an adaptive Langevin stepsize, based on the absolute value of the drift term $D^a_{x\mu} S$, in order to avoid numerical instabilities \cite{Aarts:2009dg}.
Quantum expectation values are computed as averages over the Langevin time $\theta$ after the system reaches equilibrium.

The sign problem is potentially evaded by allowing the system to explore a larger configuration space \cite{Klauder:1983nn, *Klauder:1983zm, *Klauder:1983sp,Parisi:1984cs, Aarts:2008rr,Aarts:2009uq}.
For an SU($3$) gauge theory the group is enlarged to SL($3, \mathbb{C}$).
In this context the gauge action is generalized by replacing $U^\dagger$ for $U^{-1}$, which keeps the gauge action holomorphic.

The SL($3, \mathbb{C}$) manifold is not compact, which means the system might follow a trajectory where the imaginary parts of the gauge fields are no longer small deformations compared to the real ones.
A measure of the distance from the unitary manifold is given by
\begin{align}
	d = \frac{1}{3\Omega} \sum_{x,\mu} \mathbf{Tr} \left[ U_{x\mu} U^\dagger_{x\mu} - \mathbb{1} \right] \geq 0\,,
\end{align}
with $\Omega$ being the lattice four-volume and the equality only holding if all $U_{x\mu}$ are unitary.
The \textit{gauge cooling} \cite{Seiler:2012wz} technique is employed to prevent the system from going too far from SU($3$).
It uses gauge transformations to push the system as close as possible to the unitary manifold.
In other words,
\begin{equation}
	U_{x\mu} \to e^{-\varepsilon\alpha\lambda^a f^a_x} \, U_{x\mu} \, e^{\varepsilon\alpha\lambda^a f^a_{x+\mu}}\,, \quad f^a_x = 2 \mathbf{Tr} \left[ \lambda^a \left(U_{x\mu}U^\dagger_{x\mu} - U^\dagger_{x-\mu,\mu}U_{x-\mu,\mu}\right) \right]\,,
\end{equation}
where $\alpha$, similar to $\varepsilon$, is changed adaptively based on the absolute value of $f^a_x$ \cite{Aarts:2013uxa}.
These transformations seek configurations that are closest to the SU($3$) submanifold and gauge equivalent to those generated in the Langevin process.
Gauge cooling has been shown not to affect the equilibrium probability distributions of the observables \cite{Nagata:2015uga}.

Direct simulations of full QCD \cite{Sexty:2013ica, Sinclair:2015kva} and comparisons with the hopping expansion to all orders \cite{Aarts:2014bwa} and multi-parameter reweighting \cite{Fodor:2015doa} are among recent works involving the complex Langevin method.
Discussions regarding the role of the pole arising from the logarithm of the (fermion) determinant in the complex Langevin process can be found in \cite{Splittorff:2014zca, Mollgaard:2014mga, Nishimura:2015pba}.

\section{Phase diagram of heavy-dense QCD}
The heavy-dense approximation of QCD (HDQCD) \cite{Bender:1992gn, Aarts:2008rr} is obtained by dropping the spatial hopping terms for the fermions, while the remainder can be treated exactly. The HDQCD fermion determinant reads
\begin{align}
	\det M_{HD}(U,\mu) = \prod_{N_f} \prod_{\vec{x}} \left\{ \det \left[ 1 + \left(2\kappa e^\mu \right)^{N_\tau} \mathcal{P}_{\vec{x}} \right]^2 \det \left[ 1 + \left(2\kappa e^{-\mu} \right)^{N_\tau} \mathcal{P}_{\vec{x}}^{-1} \right]^2 \right\}\,,
\end{align}
where $N_\tau$ is the temporal extent of the lattice, $N_f$ is the number of degenerate quark flavours, and the Polyakov loop and its inverse are given by
\begin{equation}
	\mathcal{P}_{\vec{x}} = \prod_{\tau = 0}^{N_\tau-1} U_{(\vec{x}, \tau), 4}\quad \text{and} \quad \mathcal{P}^{-1}_{\vec{x}} = \prod_{\tau = N_\tau-1}^{0} U^{-1}_{(\vec{x}, \tau), 4}.
\end{equation}
This model exhibits the sign and Silver Blaze \cite{Cohen:2003kd} problems, but is computationally cheaper to simulate than full QCD.
Other studies include combinations with strong-coupling expansions \cite{deForcrand:2014tha,Glesaaen:2015vtp}, reweighting \cite{DePietri:2007ak}, histogram \cite{Ejiri:2013lia}, density of states methods \cite{Garron:2016noc} and mean-field methods \cite{Rindlisbacher:2015pea}.
Other complex Langevin studies of this model have led to important algorithmic improvements \cite{Aarts:2008rr,Aarts:2009dg,Seiler:2012wz,Aarts:2014bwa}.

Our study consisted of an extensive scan of the $T-\mu$ plane with a total of $880$ ensembles using different combinations of $N_\tau$ and $\mu$ for each of the three different lattice volumes, described in Table \ref{tb.simulation.parameters}.
The critical chemical potential $\mu^0_c = -\ln(2 \kappa)$ indicates the transition to higher densities at zero temperature ($N_\tau \to \infty$).
We have used fixed gauge coupling $\beta = 5.8$ and hopping parameter $\kappa=0.04$, resulting in an approximate lattice spacing of $a \sim 0.15$ fm, which has been determined using the gradient flow \cite{Borsanyi:2012zs}.

From the analysis of the distance from the SU($3$) manifold and distributions of observables we have learned that the complex Langevin method combined with gauge cooling produces correct results as long as $d \lesssim \mathcal{O}(0.1)$ \cite{Aarts:2016qrv}.
Therefore we present here simulation results for which the distance is less than $0.03$.
\begin{table}[h]
		\begin{tabular}{ccccccccccccccc}
				\hline
				\multicolumn{5}{c|}{$\beta = 5.8$} & \multicolumn{5}{c|}{$V = 6^3, 8^3,
				10^3$} & \multicolumn{5}{c}{$a\sim 0.15$ fm} \\
				\multicolumn{5}{c|}{$\kappa = 0.04$} & \multicolumn{5}{c|}{$N_f = 2$} &
				\multicolumn{5}{c}{$\mu_c^0=2.53$} \\
				\hline
				$N_\tau$ & 28 & 24 & 20 & 16 & 14 & 12 & 10 & 8 & 7 & 6 & 5 & 4 & 3 & 2 \\
				$T$ [MeV] & 48 & 56 & 67 & 84 & 96 & 112 & 134 & 168 & 192 & 224 & 268 &
				336 & 447 & 671\\  
				\hline
		\end{tabular}
		\caption{\label{tb.simulation.parameters}  
		Simulation parameters used in this study. The chemical potential $\mu$ has been varied from 0 to $1.3\,\mu_c^0$, with $\mu_c^0=-\ln(2\kappa)$.
		The (approximate) lattice spacing has been set using the gradient flow~\protect\cite{Borsanyi:2012zs}.
		}
\end{table}

The main observable we have used in this work is the average of the traced (inverse) Polyakov loop, defined by
\begin{align}
	\langle P \rangle = \frac{1}{V} \sum_{\vec{x}} \langle P_{\vec{x}} \rangle\, \quad &\text{with} \quad P_{\vec{x}} = \frac{1}{3} \mathbf{Tr} \mathcal{P}_{\vec{x}}\,,\\
	\langle P^{-1} \rangle = \frac{1}{V} \sum_{\vec{x}} \langle P^{-1}_{\vec{x}} \rangle\, \quad &\text{with} \quad P^{-1}_{\vec{x}} = \frac{1}{3} \mathbf{Tr} \mathcal{P}^{-1}_{\vec{x}}\,.
\end{align}
It is worth noting that $P_{\vec{x}}$ and $P^{-1}_{\vec{x}}$ are complex-valued but their expectation values are real, as they are proportional to $e^{-F}$ with $F$ being the free energy of a single (anti) quark.
We consider the symmetrised combination, which is real for SU($3$) gauge configurations,
\begin{equation}
	P^s_{\vec{x}} = \frac{1}{2} \left( P_{\vec{x}} + P^{-1}_{\vec{x}} \right)\,.
\end{equation}
In \cite{Aarts:2016qrv} results for the density and its susceptibility can be found as well.

\begin{figure}[h!]\center
	\includegraphics[scale=0.46]{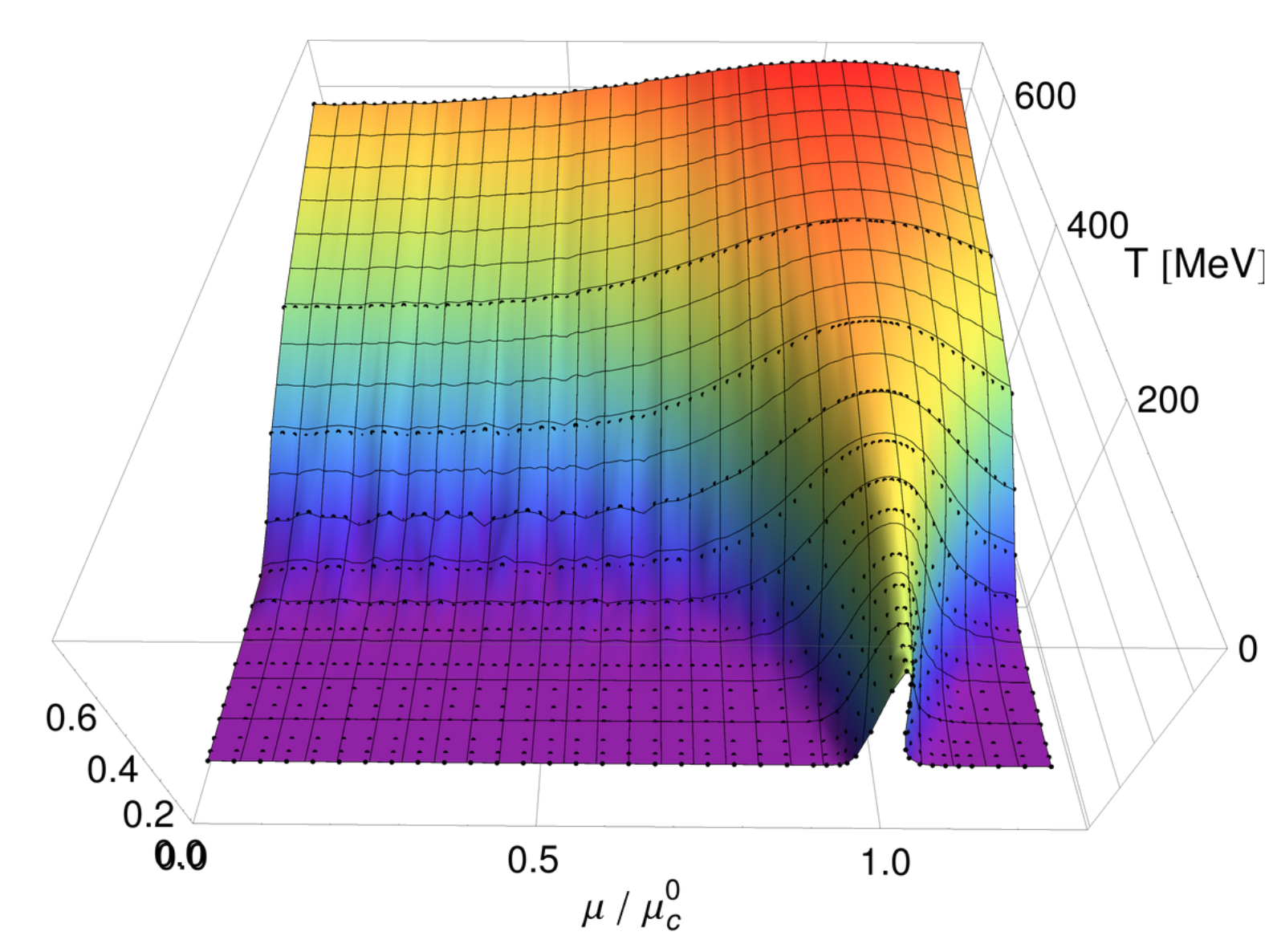}
	\includegraphics[scale=0.46]{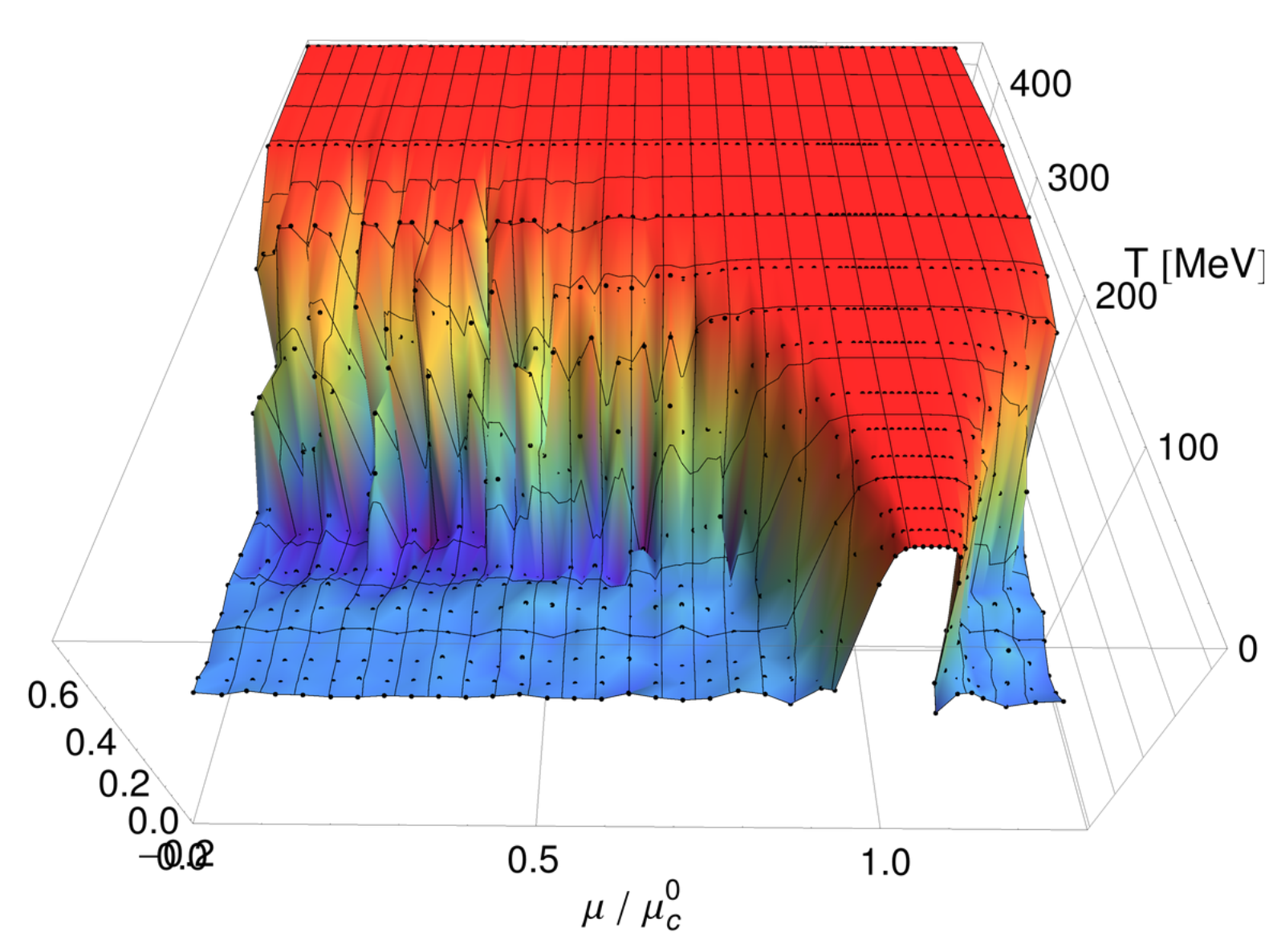}
	\caption{\label{fig.3d.plots}Average value (left) and Binder cumulant (right) for the symmetrised Polyakov loop as functions of $T$ and $\mu$ for $V=10^3$. Each black dot indicates a simulation point and the interpolated surface is added to guide the eye.}
\end{figure}

The left panel of Figure \ref{fig.3d.plots} shows a $3$D plot of the average symmetrised Polyakov loop as function of $T$ and $\mu$ for a volume of $10^3$ and the parameters given in Table \ref{tb.simulation.parameters}. Each black point represents the average from an individual simulation and the cubic spline surface is plotted to guide the eye.
The Polyakov loop exhibits both the transition to high densities, which is induced by the quark dynamics, and the thermal deconfinement transition, driven by gluonic dynamics.

For the determination of the boundary between the different phases we have studied the Binder cumulant \cite{Binder:1981sa} of $P^s$,
\begin{equation}
	B = 1 - \frac{\langle (P^s)^4 \rangle}{3 \langle (P^s)^2 \rangle^2}\,.
\end{equation}
The result is shown on the right hand side of Figure \ref{fig.3d.plots}, where a clear separation between the confined, with $B=0$, and deconfined, $B=2/3$, phases is visible.

\begin{figure}[h!]\center
	\includegraphics[scale=0.70]{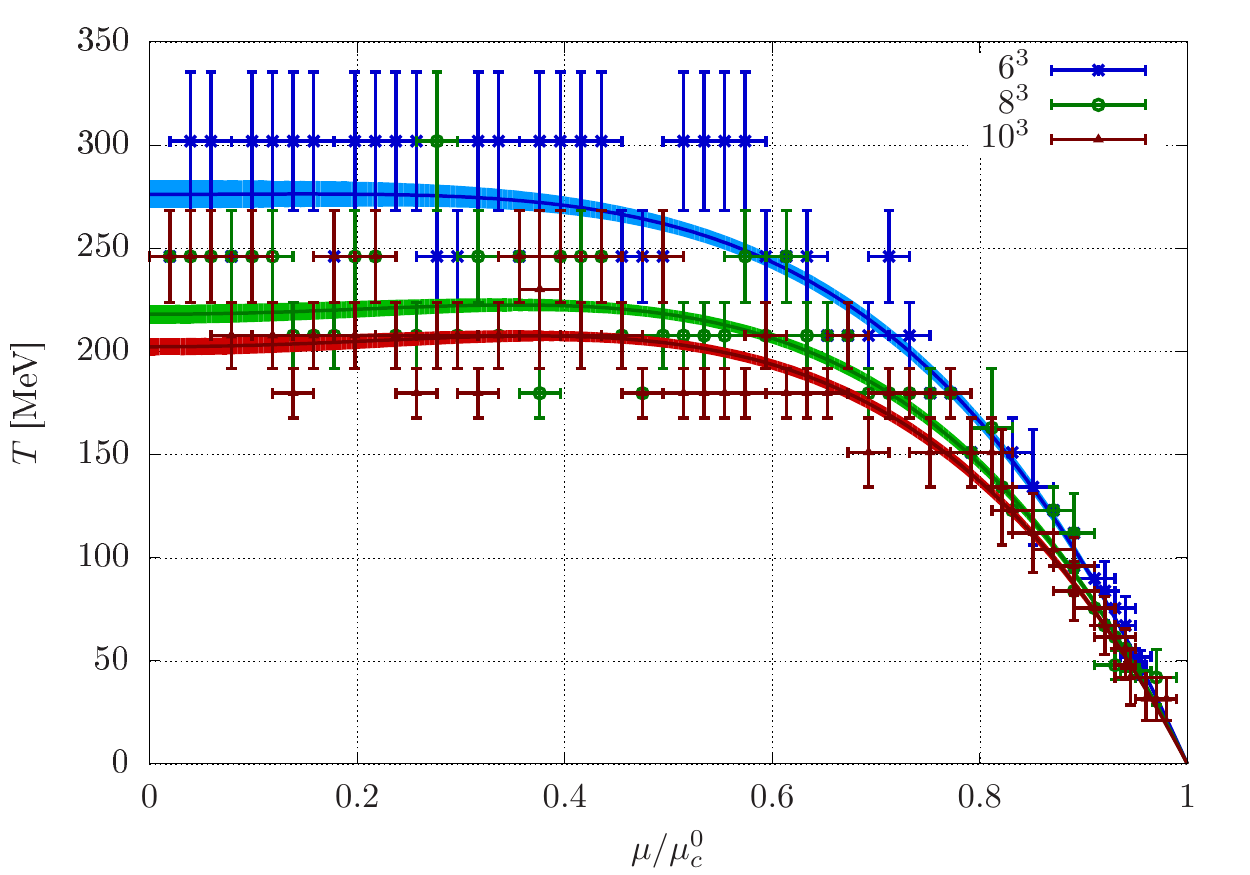}
	\caption{\label{fig.fit.phase.diagram}Phase boundary of HDQCD for different volumes, together with polynomial fits.}
\end{figure}

Figure \ref{fig.fit.phase.diagram} shows the deconfinement transition in the $T-\mu$ plane, along with a polynomial fit, for the lattice volumes of $6^3$, $8^3$ and $10^3$.
The points and error bars represent the region where the Binder cumulant increases from $0$ to $2/3$.

Finite size effects are clearly visible in Figure \ref{fig.fit.phase.diagram}, but are less pronounced for the two larger volumes.
The polynomials shown are of the form $T_c(\mu) = b_1 [1-(\mu/\mu^0_c)^2] + b_2 [1-(\mu/\mu^0_c)^2]^2$, which enforces that $T_c(\mu^0_c) = 0$.
Comparisons with higher order polynomials have revealed that this is sufficient for our data set.
A study of different functions, including an expected non-analiticity at $\mu = \mu^0_c$, along with the fitted values for the polynomials can be found in \cite{Aarts:2016qrv}.

In principle, the Binder cumulant can be used to determine the order of the phase transition, since at the transition point its value depends only on the universality class \cite{Binder:1981sa}.
However, due to our choice of fixed lattice spacing our values for the temperature were constrained by $N_\tau$ being an integer number.
A deeper analysis of the volume dependence would require a more precise determination of the critical temperature $T_c$ as function of $\mu$, with smaller error bars.

\section{Instabilities}
During our analysis we have found cases where after a certain Langevin time the distribution of the observables would change, typically becoming wider or shifting their average values.
Based on the formal justification \cite{Aarts:2009uq,Aarts:2011ax} and comparisons with reweighting \cite{DePietri:2007ak} we concluded that these instabilities do not reflect the original theory.

We have found evidence that a large distance from SU($3$) is correlated with the appearance of the instabilities.
Therefore, we have imposed a conservative cutoff on the distance, i.e., exclude data points where $d > 0.03$.
A more detailed discussion is found in \cite{Aarts:2016qrv}.

\section{Conclusion}
We have presented here our results on the phase diagram of QCD in the limit of heavy quarks, using complex Langevin simulations.
The combination with gauge cooling allowed us to perform ab-initio simulations on the whole $T-\mu$ plane.
We have mapped the phase boundary among different phases using the Binder cumulant for the symmetrised Polyakov loop and described the boundary line by a polynomial.

Despite some instabilities during the Langevin process, which happen when the system is too far from the unitary submanifold even after gauge cooling is applied, we have been able to collect enough data for a reliable analysis.

The prospect for the future would be to improve the precision in the determination of the phase boundary, and also the order of the transition.
For that it will be necessary to change the temporal extent and the lattice spacing simultaneously.
Furthermore, a better control of the Langevin process is important and work in this direction is under development.
\acknowledgments

We thank Erhard Seiler and Ion-Olimpiu Stamatescu for collaboration on related topics. 
We are grateful for the computing resources made
available by HPC Wales.
We acknowledge
the STFC grants ST/L000369/1,                                                         
ST/K000411/1,  ST/H008845/1, ST/K005804/1 and ST/K005790/1,     
the STFC DiRAC HPC Facility (www.dirac.ac.uk),
STFC grant ST/L000369/1, the
Royal Society and the Wolfson Foundation. FA is grateful for the support
through the Brazilian government programme ``Science without Borders'' under scholarship number
BEX 9463/13-5.

\bibliographystyle{JHEP}
\bibliography{././bib/grilo,././bib/papers,././bib/proceedings}{}

\end{document}